\documentclass{ifacconf}

\usepackage{graphicx}      % include this line if your document contains figures
\usepackage{natbib}        % required for bibliography
\usepackage{amssymb,amsmath}
\usepackage{epstopdf}

\usepackage[dvipsnames]{xcolor}

\renewcommand\epsilon{\varepsilon}
\begin{document}
\begin{frontmatter}

\title{Gamma Rhythm Analysis and Simulation Using Neuron Models} 
% Title, preferably not more than 10 words.

\thanks[footnoteinfo]{
Data analysis (Section~\ref{sec:data}) was supported by the grant of Saint Petersburg State University (No. 84912397).
\\
Model synthesis for simulation of gamma rhythm (Section~\ref{sec:sym}) was performed in Lobachevsky State University of Nizhny Novgorod and supported by Russian Science Foundation (project No. 19-72-10128).}

\author[First,Second]{Evgeniia S. Sevasteeva}
\author[Second,Third]{Sergei A. Plotnikov} 
\author[Fourth]{Dmitry R. Belov}

\address[First]{Saint Petersburg State University, Universitetskii Pr. 28, St. Petersburg 198504, Russia (e-mail: esevasteeva@outlook.com)}
\address[Second]{Institute for Problems of Mechanical Engineering, Russian Academy of Sciences, Bolshoy~Ave 61, Vasilievsky Ostrov, St.~Petersburg, 199178, Russia  (e-mail: waterwalf@gmail.com).}
\address[Third]{Lobachevsky State University of Nizhny Novgorod, Gagarina Pr. 23, Nizhny Novgorod, 603950, Russia}
\address[Fourth]{Almazov National Medical Research Centre, Akkuratova St. 2, St.~Petersburg 197341, Russia (e-mail: dmbelov64@mail.ru).}

\begin{abstract}    
 Neural oscillations are electrical activities of the brain measurable at different frequencies. This paper studies the interaction between the fast and slow processes in the brain. We recorded signals intracranially from the simple Wistar rats, performed the signal processing, and computed the correlation between envelopes of the high-frequency gamma rhythm and a low-frequency ECoG signal. The analysis shows that the low-frequency signal (delta rhythm) modulates the gamma rhythm with a small time delay. Further, we used simple excitable neuron models, namely  FitzHugh-Nagumo and Hindmarsh-Rose, to simulate the gamma rhythm. The low-frequency signal delta rhythm can be used as the input to affect the threshold and simulate gamma rhythm using these neuron models.
\end{abstract}


\end{frontmatter}
%===============================================================================

\section{Introduction}
Neural oscillations may represent variable signals underlying flexible communication within and between cortical areas.
 Neural oscillations are electrical activities of the brain measurable at different frequencies. They are typically described as low-frequency bands at delta ($<4$~Hz), theta ($4-8$~Hz), alpha ($8-12$~Hz), and beta ($12-30$~Hz) to high-frequencies at gamma band that spans from roughly gamma ($30-80$~Hz) to high gamma (HG) ($80-150$~Hz) \cite{Moran2011}. These oscillatory activities can be obtained at many levels, ranging from individual neurons to large-scale synchronized interactions between neurons,  which can be observed by an electroencephalogram, intracranial electrical recordings, or magnetoencephalography and characterised by different frequency, amplitude and phase. The interactions between oscillations of different frequencies have been shown previously by \cite{Jaime2017, Jackson2011}. Particularly, the coupling between low-frequency and high-frequency bands have been studied in \cite{JEN07, Canolty2006}.
 
 %These oscillatory activities can be obtained at many levels, ranging from single cell to local field potentials in animals, to large-scale synchronized activities in human scalp. Neuronal oscillations in different frequency bands have been reported in many studies. On the other hand, there are some studies of the coupling between low-frequency and high-frequency bands, see \cite{JEN07a} and references therein.  
\cite{Canolty2006} showed that the phase of the low-frequency theta rhythm modulates the amplitude of the high gamma band of the electrocorticogram. %with stronger modulation occurring at higher theta amplitudes. 
It was shown that HG power is modulated by theta phase and an increase in theta power strengthens theta/HG coupling.
The co-expression of theta with two gamma frequency generators in the subiculum obviously leading to hippocampal communication with other brain regions was shown in \cite{Jackson2011}.

%Note, that in \cite{Canolty2006} it was observed coupling between the high- and low-frequency bands of neural oscillations in the human brain. The phase of the low-frequency theta rhythm modulates power in the high gamma band of the electrocorticogram, with stronger modulation occurring at higher theta amplitudes. 

In \cite{Moran2011} the concept that oscillation abnormalities in gamma band in schizophrenia often occur in the background of oscillation abnormalities of low-frequency bands was discussed.
\cite{Neske2016} gathered the information on the mechanisms and functions of the slow ($<1 $Hz) oscillation in the cortex and thalamus, characterising their spontaneous activity, produces by  neurons fluctuation between periods of intense synaptic activity (Up states) and almost complete
silence (Down states). It was noticed there that phase-amplitude coupling could provide a mechanism for spatial and temporal neural information processing in a hierarchical manner. 
Grouping of higher-frequency oscillations by the slow oscillation  is the process continuously taking place in the brain as oscillations of one
frequency band are rarely expressed alone. 
%It was noted in \cite{Neske2016} that during even the most inactive periods, the cortex and thalamus express rich spontaneous activity in the form of slow ($<1 $Hz), synchronous network state transitions. Throughout this so-called slow oscillation, cortical and thalamic neurons fluctuate between periods of intense synaptic activity (Up states) and almost complete silence (Down states). The two decades since the original characterization of the slow oscillation in the cortex and thalamus have seen considerable advances in deciphering the cellular and network mechanisms associated with this prevalent phenomenon \cite{Neske2016}.
%These oscillations are generated by large ensembles of neurons that oscillate synchronously and provide the basis for complex flowing cognitive processes (memory, thinking, planning and decision-making) \cite{ENG01,JEN07a,BUZ10}.  

Nevertheless, there is still an open question of how  the fast and slow processes in the brain interact with each other. This interaction can be achieved in different ways: by means of amplitude correlations \cite{BRU04,PAL05}, phase to phase synchronization \cite{PAL05}, phase to frequency principle \cite{JEN07} or phase to power locking \cite{BRU04,Canolty2006,OSI08}. 

Power to power coupling is particularly interesting since the high-frequency rhythm (e.g. gamma) can be generated by a low-frequency signal (e.g. delta, theta or alpha). This fact can be used for the simulation of the excitable system producing high-frequency oscillations. This paper continues the research presented earlier at the conferences Physcon 2019 and DCNA 2021 \cite{PLO19,SEV21}.

The rest of this paper is organized as follows: In Sec.~\ref{sec:data}, the signal processing and data analysis of ECoG rhythms are performed. In Sec.~\ref{sec:sym}, the gamma rhythm using the neuron models with disturbances and low-frequency signal as input is simulated. Subsec.~\ref{sec:FHN} considers the FitzHugh-Nagumo model, while Subsec.~\ref{sec:HR} deals with the Hindmarsh-Rose model. %Discussion is given in Sec~\ref{sec:disc}, while
Conclusion is drawn in Sec.~\ref{sec:concl}.

\section{Signal Recording}
The ECoG recording of the simple Wistar rats was carried out in the course of acute experiments under anesthesia. The recordings of low-resistance electrodes usually contain mostly slow dynamics and almost do not have high frequency rhythms, since high frequency signals are bounded by the small areas, while the synchronous activity of vast areas of the brain is needed for the low frequency signals generation \cite{CSI03,FUR13}. Therefore the Neuronexus E$32-600-10-100$ multi-electrode array with $32$ registration sites of $100$ $\mu$m each (site impedance is $500$~k$\Omega$) with cross-site $600$ $\mu$m intervals was used for the gamma rhythm recording.
These characteristics give a stable gamma rhythm ($30-80$ Hz) recording on a small scale with sufficient locality, and even under general anesthesia. This multi-electrode array was placed in the left hemisphere approximately in the area of sensorimotor cortex.
To collect the low frequency signal the gold plated screw electrodes (impedance is $25-50$~k$\Omega$) were used. One such screw was also used as an indifferent electrode over the cerebellum. As expected low-resistance electrodes gave a slow rhythm under anesthesia, i.e. showed the usual low-frequency oscillations typical for sleep and anesthesia.

Three trials with an approximate duration of $120$ seconds, with the sample rate of $2$ kHz were made on different rats. For the analysis we consider the fragment of one recording without instrumental artifacts identified by visual inspection with a duration of $30$ seconds. The recordings of two electrodes were chosen under consideration: one high-resistance and one low-resistance electrode.  Figure~\ref{fig1}~(a) presents the unfiltered fragment of the ECoG recording of two sites. The recording of the high-resistance electrode is marked by blue color, while the recording of the low-resistance electrode is marked by red color. The recording of the high-resistance electrode will be used to get gamma rhythm, while the recording of the low-resistance electrode will be used to get slow delta rhythm. 

\section{Data Analysis}\label{sec:data}
This section studies the interdependence between fast and slow processes in the brain.  There are different principles of cross-frequency interactions, e.g. phase-locking between oscillations at different frequencies \cite{OSI08}; phase to frequency principle \cite{JEN07}; phase to power modulation \cite{Canolty2006}; and power to power correlation \cite{BRU04}. Here we use the signal-envelope correlation as an alternative measure that can detect coupling between gamma and low-frequency rhythms. It differs from the approach of amplitude-envelope correlation proposed in \cite{BRU00}, since it posts that oscillations in power of the faster gamma oscillations are correlated with power changes in the lower frequency band. We will show that the faster gamma oscillations arise when the value of low-frequency rhythm is higher than some threshold value, and on the other hand, there is gamma oscillation death when the value of low-frequency rhythm is lower than this threshold.
\subsection{Signal Processing}
\begin{figure}
\flushleft
\begin{minipage}[h]{1\linewidth}
\center{\includegraphics[width=1\linewidth]{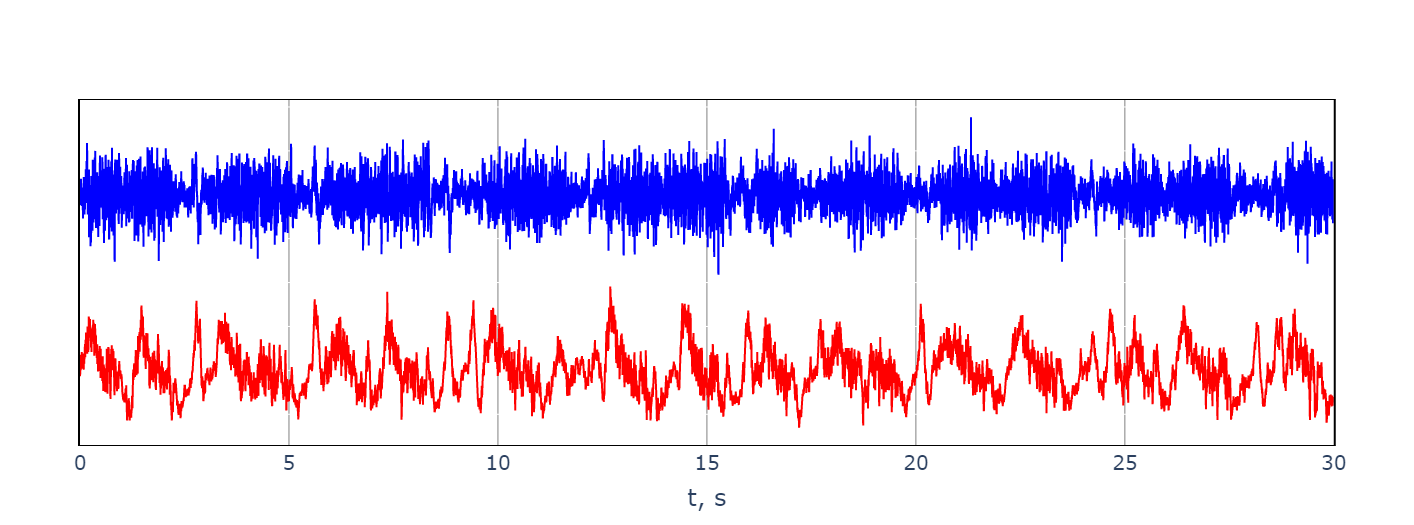} \\ (a)}
\end{minipage}
\vfill
\begin{minipage}[h]{1\linewidth}
\center{\includegraphics[width=1\linewidth]{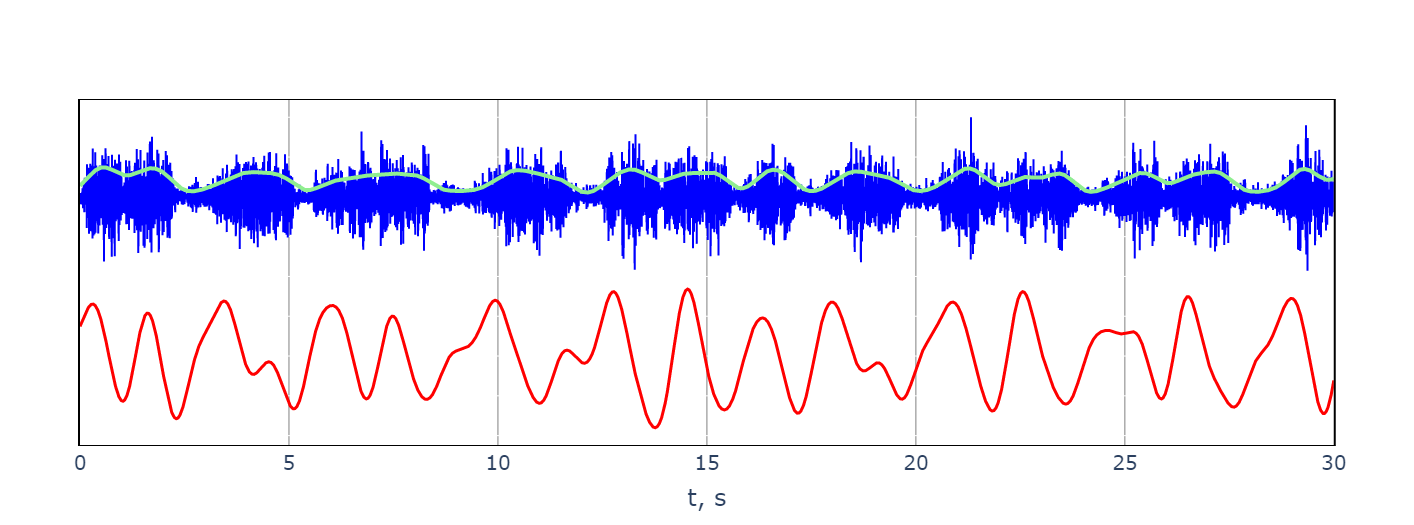} \\ (b)}
\end{minipage}
\caption{The fragments of the ECoG recordings. The blue color marks the high-resistance electrode recording, namely the high-frequency gamma rhythm $x(t)$, while the green color marks its envelope $e(t)$. The red color marks the low-resistance electrode recording, namely low-frequency delta rhythm $y(t)$.}
\label{fig1}
\end{figure}

The first step of analysis is to determine the signals within an appropriate frequency band. Both variants with finite impulse response (FIR) and infinite impulse response (IIR) were considered as filters. The use of the first type filters was abandoned later due to the fact that the order of such filters is too high to obtain the desired band characteristics \cite{MIT11}. Therefore, the Butterworth filter \cite{RAB75}  was chosen as the IIR filter with the smoothest amplitude-frequency response at passband frequencies. To avoid introducing nonlinear phase shifts that are critical in ECoG signals analysis, a zero-shift filter was implemented \cite{MIT11}, which means that the Butterworth filtering was applied to the data in the forward and backward directions. The recording of the high-resistance electrode is digitally bandpass-filtered in the frequency domain between $30$ and $80$ Hz to obtain a gamma rhythm. This filtered fast-frequency signal will be denoted as $x(t)$. One can see in Fig.~\ref{fig1}~(b) that there are periods of gamma activation and extinction of approximately $1$ second duration, i.e. the whole time period approximately equals $2$ seconds. Hence, we choose the frequency domain for the low-resistance electrode recording between $0$ and $0.5$ Hz to show the correlation between this low-frequency signal and gamma activation and extinction. We denote this filtered low-frequency signal as $y(t)$.

Figure~\ref{fig1}~(b) presents the two filtered signals. One can see that there is interconnection between the low-frequency signal $y(t)$ and gamma rhythm $x(t)$. The increasing of low-frequency rhythm leads to the gamma rhythm activation, while its decreasing leads to the gamma rhythm extinction. This means that there is the threshold value of low-frequency rhythm. The value of $y(t)$ higher than the threshold leads to gamma activation, while the value of $y(t)$ lower than the threshold leads to gamma extinction. To find the numerical measure of two signals interdependence we should somehow transform the high-frequency signal $x(t)$. It characterizes by the alternating intervals of activation and extinction of oscillations. Therefore, one can calculate the upper envelope which will go around these intervals of activation and extinction. Since we choose the frequency domain for the low-resistance signals between $0$ and $0.5$ Hz (i.e. we identify low-frequency signal as a delta rhythm), the envelope frequency of the gamma rhythm will be the same. There are different ways to calculate the envelope. The following methods were considered: peak-envelope, empirical mode decomposition (EMD) \cite{GUP19} and Hilbert transform \cite{PAN65}. First two methods show good results but it is necessary to select a suitable reference point (peak-envelope) or number of components for envelope (EMD) manually for each experiment. The Hilbert transform also works well and does not need for individual customization. However, it provides too detailed envelope. Since that the low-pass zero-phase filter was implemented to the Hilbert transform envelope. Denote the envelope of the gamma rhythm by $e(t)$. The result of calculating the envelope is presented in Fig.~\ref{fig1}~(b) and is marked by green color.

Now we find the mean value of two signals:
\begin{equation}\label{f1}
\bar y=\frac{1}{|T|}\sum_{t\in T}y(t),\quad \bar e=\frac{1}{|T|}\sum_{t\in T}e(t),
\end{equation}
where $T$ is a set of all signal measurements, and $|T|$ is its cardinality, since these signals are discrete; and center of the signals are: 
\begin{equation}\label{f2}
y_c(t)=y(t)-\bar y,\quad e_c(t)=e(t)-\bar e.
\end{equation}
Moreover, we normalize the obtained signals:
\begin{equation}\label{f3}
y_n(t)=\frac{1}{\max\limits_{t\in T}|y_c(t)|}y_c(t),\quad e_n(t)=\frac{1}{\max\limits_{t\in T}|e_c(t)|}e_c(t).
\end{equation}

\begin{figure}
\center{\includegraphics[width=1\linewidth]{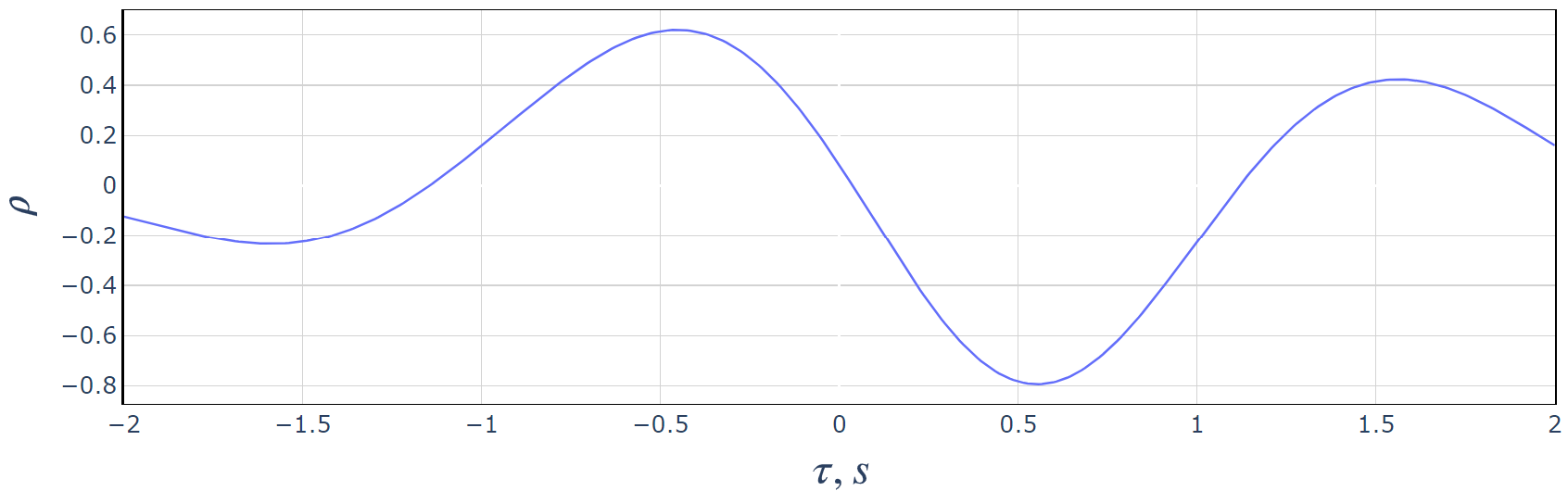}}
\caption{Dependence of the cross-corellation $\rho$ between the low-frequency signal $y_n(t+\tau)$ and the high-frequency signal envelope $e_n(t)$ on the delay $\tau$ between two signals.}
\label{fig3}
\end{figure}

Thus, the values of transformed signals now belong to the interval $[-1;1]$. Now we are ready to find the numerical measure of signals interdependence. For this purpose we use Pearson correlation coefficient, which is a statistic measure of linear correlation between two signals, and it has a value between $-1$ and $1$. Since there might be time delays between two signals, we calculate the correlation coefficient for different time delays:
\begin{equation}\label{f4}
\rho(\tau)=\frac{\sum\limits_{t\in \hat T}y_n(t+\tau)e_n(t)}{\sqrt{\sum\limits_{t\in \hat T}y_n^2(t+\tau)\sum\limits_{t\in \hat T}e_n^2(t)}},
\end{equation}
where $\tau$ is a delay, $\hat T$ is a set, which contains all elements of $T$ except for first $4000$ and last $4000$, since the sample rate of the signal is equal to $2$ kHz and the frequency of considered signals is $0.5$ Hz, i.e. the possible delay $\tau$ belongs to the interval $[-2;2]$ seconds. Figure~\ref{fig3} presents the dependence of cross-correlation $\rho$ on the delay $\tau$ between two signals. The maximum value of cross-correlation is $0.62$ which corresponds to the delay $\tau=-0.456$ seconds. This means that slow delta rhythm modulates the high-frequency gamma rhythm after time delay equals $456$ ms. One more argument for this statement is that the correlation curve has explicit negative minima for positive $\tau = 0.566$, i.e. two researched signals enter in the opposite phase after about a half of the slow signal $y(t)$ period. Note that the period of function $\rho(\tau)$ is approximately $2$ seconds, which is the same as the period of low-frequency delta rhythm $y(t)$.

\begin{figure}
\center{\includegraphics[width=1\linewidth]{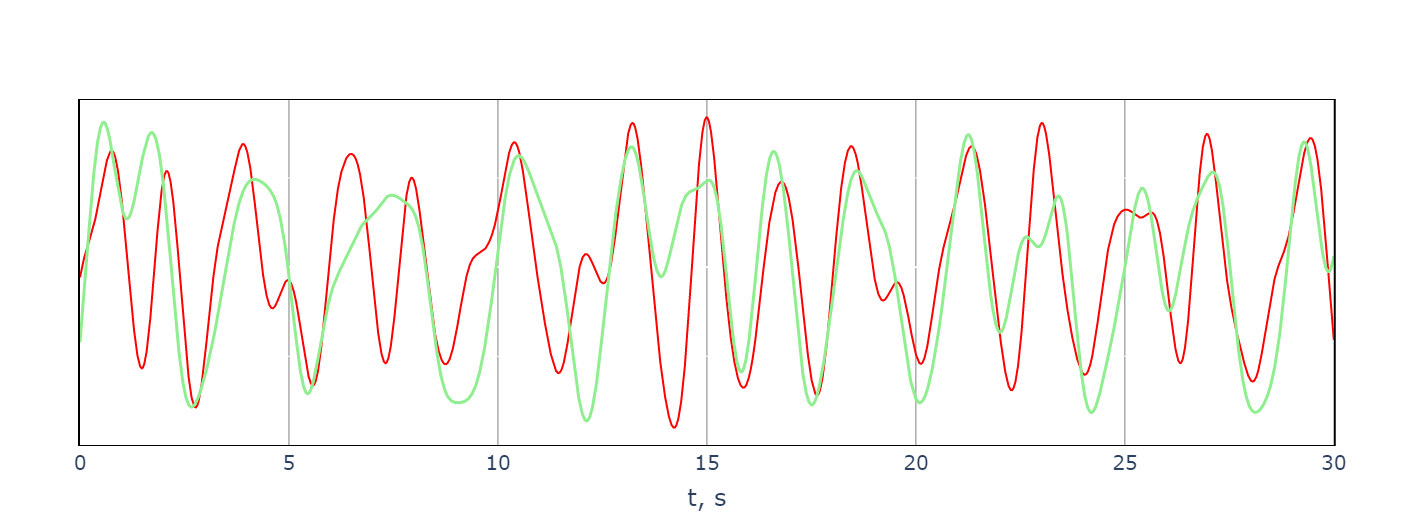}}
\caption{The relationship between two centered and normalized signals: the envelope of gamma rhythm $e_n(t)$ marked by green color and the shifted low-frequency delta rhythm $y_n(t-0.456)$ marked by red color.}
\label{fig4}
\end{figure}

Figure~\ref{fig4} presents the dynamics of two processed signals, namely centered and normalized envelope of gamma rhythm $e_n(t)$ marked by the green color and shifted centered and normalized delta rhythm $y_n(t-0.456)$ marked by red color. One can see that there is dependence of the increasing of the envelope value, i.e. the arising of gamma oscillations, on the increasing of delta rhythm value.

\section{Simulation of Gamma Rhythm}\label{sec:sym}
A strongly coupled group of active neurons (cell ensemble) is a single cognitive unit, and the spike activity of such a group is the basic unit of neural coding. The internal mechanisms of regulation of these ensembles provide the basis for complex flowing cognitive processes (memory, thinking, planning and decision-making) \cite{BUZ10}. These ensembles of neurons oscillating in synchrony produce the neuronal oscillations, i.e. EEG rhythms \cite{TRA97}. The dynamics of a single neuron can be described by the differential equations of the excitable system, for instance, FitzHugh-Nagumo (FHN) \cite{FIT61a,NAG62} and Hindmarsh-Rose \cite{HIN84} (HR) models.

\subsection{FitzHugh-Nagumo model}\label{sec:FHN}
The FHN model is described by two differential equations with cubic nonlinearity:
\begin{equation}\label{f5}
\begin{aligned}
\epsilon\dot u(t)&=u(t)-\frac{u^3(t)}{3}-v(t) + I, \\
\dot v(t)&=u(t)+a - b v(t),
\end{aligned}
\end{equation}
where $u$ and $v$ represent the state variables of a neuron meaning the membrane potential and the recovery variable, respectively. $I$ is an external stimulus. $0<\epsilon < 1$ separates the fast and slow dynamics; $a$ is the threshold parameter: for $ a > 1 - 2b/3$ the system is excitable, i.e. it has a locally stable equilibrium point on the phase plane, while for $ 0 < a < 1 - 2b/3$ it is oscillatory, i.e. it has a stable limit cycle on the phase plane.

Choose $\epsilon=0.8$, which is a value for neuron dynamics. For $a=0.3,  ~b = 0.8$ the system \eqref{f5} is in oscillating regime, while the period of oscillation is equal to $3.2$ seconds, which differs from the oscillation period of gamma rhythm (neural oscillation with a frequency between $35$ and $80$ Hz). To make it the same one can change the time, i.e. introduce new time $\tilde t=\delta t$, i.e. $\delta$ is a time scaling coefficient. Then the system \eqref{f5} equation can be rewritten as (with an omitted tilde):
\begin{equation}\label{f6}
\begin{aligned}
\dot u(t)&=\frac{\delta}{\epsilon}\left[ u(t)-\frac{u^3(t)}{3}-v(t) +I\right], \\
\dot v(t)&=\delta\left[u(t)+a - b v(t)\right].
\end{aligned}
\end{equation}
Choosing $\delta$ big enough one can make the system \eqref{f6} solution oscillate with the same frequency as gamma rhythm $x(t)$. For this purpose we choose $\delta=325$. 

We showed that the gamma rhythm activation depends on the value of delta rhythm $y(t)$ with a delay $\tau$. Then we add this signal to the first equation of \eqref{f6} as an external stimulus to affect the threshold $a$. Moreover, we add the ``neuronal'' noise $\xi(t)$ which is assumed to be an unbiased Gaussian white noise which is a mathematical description of many natural processes. Thus, with these additions the system \eqref{f6} can be presented in the form:

\begin{equation}\label{f7}
\begin{aligned}
\dot u(t)&=\frac{\delta}{\epsilon} \left[ u(t)-\frac{u^3(t)}{3}-v(t)+\xi(t) + y(t+\tau)\right], \\
\dot v(t)&=\delta\left[u(t)+a- b v(t)\right].
\end{aligned}
\end{equation}

\begin{figure}
\flushleft
\begin{minipage}[h]{1\linewidth}
\center{\includegraphics[width=1\linewidth]{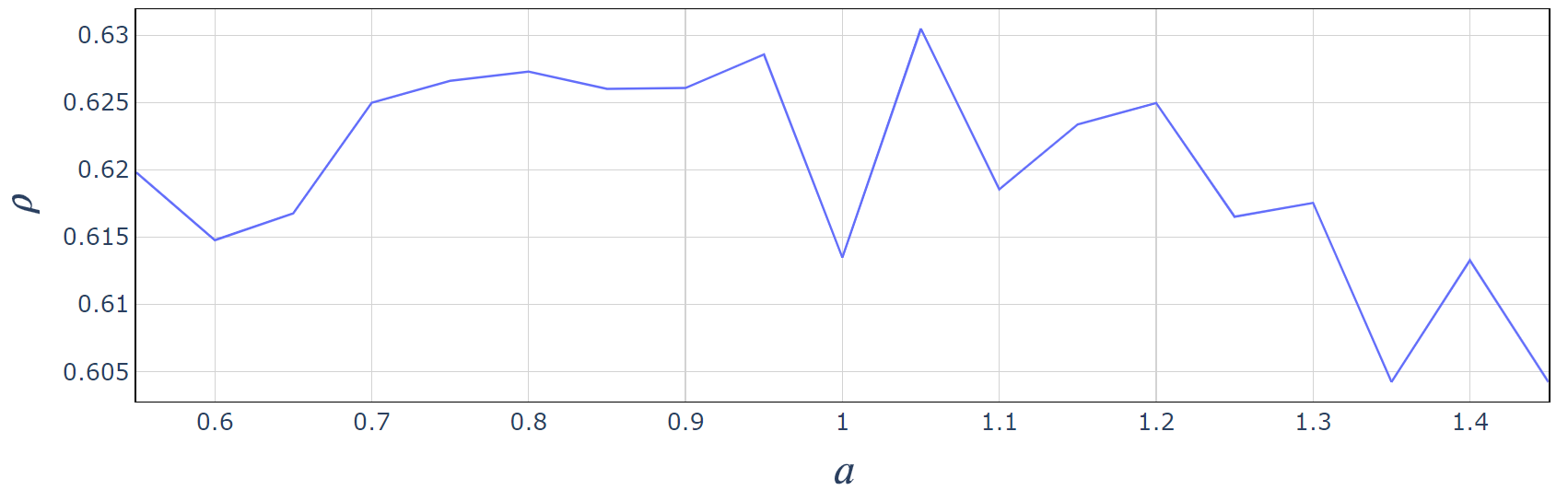} \\ (a)}
\end{minipage}
\vfill
\begin{minipage}[h]{1\linewidth}
\center{\includegraphics[width=1\linewidth]{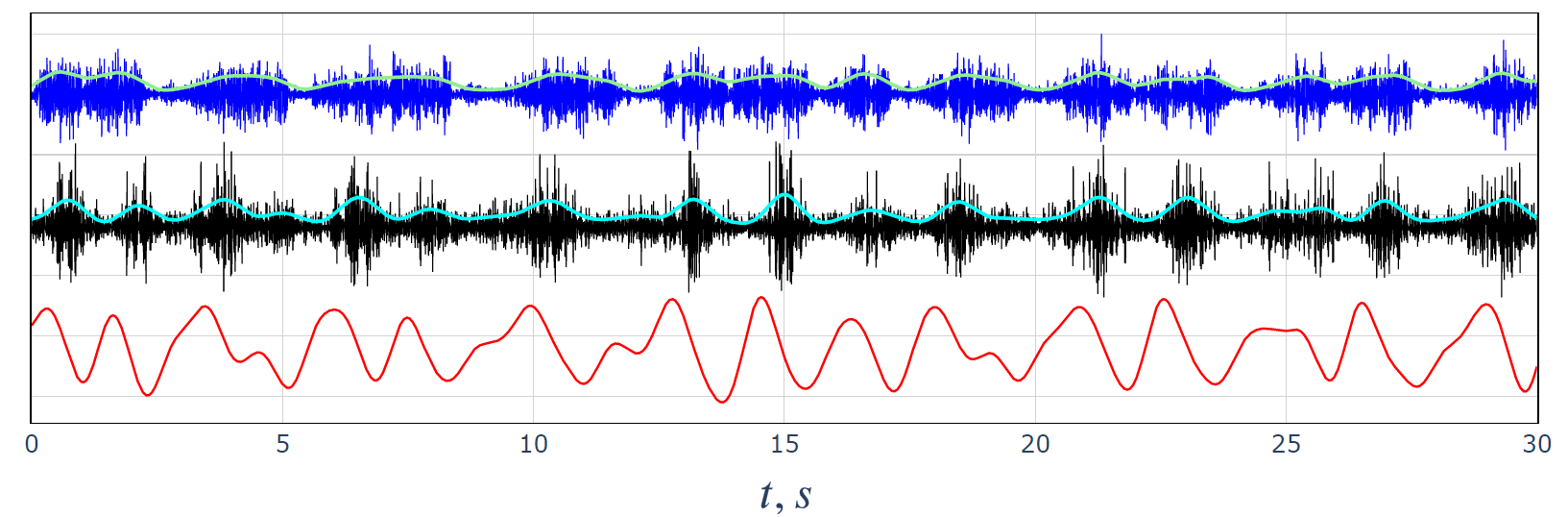} \\ (b)}
\end{minipage}
\caption{The comparison of experimental and generated by FHN model signals. (a): experimental $e_n$ and FHN model $e_{sim}$ envelopes correlation dependence on the parameter $a$; (b): dynamics of the FHN system membrane potential $u(t)$ \eqref{f7} (marked by black color) and its envelope (marked by cyan color) with comparison to the high-frequency gamma rhythm $x(t)$ (marked by blue color) and its envelope (marked by green color). The red color marks the low-frequency delta rhythm $y(t)$. System parameters: $\epsilon=0.8$, $\delta=325$, $a=1.05$, $b=0.8$, $\tau=-0.456$. The initial conditions are zeros.}
\label{fig5}
\end{figure}

The last thing we need to do is to find the appropriate value of the threshold $a$. For this purpose we vary the value of the threshold $a$ between $0.55$ and $1.4$ with a step $0.05$ - with these values initial system \eqref{f6} has stable equilibrium point. For every value of the threshold $a$ we simulate the dynamics of the system \eqref{f7} and calculate its normalized and centered envelope $e_{sim}$. For this purpose, we use the Pearson correlation coefficient, which is a statistical measure of linear correlation between high-frequency signal envelope $e_n$ and simulated signal envelope $e_{sim}$. Result are presented on the Fig.~\ref{fig5}~(a). The maximal value of correlation is $0.63$, which corresponds to the threshold $a=1.05$.

Figure~\ref{fig5}~(b) presents the results of the simulation. One can see that the low-frequency signal $y(t)$ affects the FHN system \eqref{f7} threshold: the system exhibits self-sustained periodic firing, when $y(t)$ value becomes high enough, while there is the oscillation death when $y(t)$ value is low. The periods of FHN system \eqref{f7} firing coincide with the periods of gamma rhythm $x(t)$ activation. Thus, we can use the FHN model to simulate the gamma rhythm.

\subsection{Hindmarsh-Rose model}\label{sec:HR}
The HR model is described by three nonlinear differential equations:
\begin{equation}\label{f8}
\begin{aligned}
\dot u(t)&=v(t)-a u^3(t)+bu^2(t)-w(t) + I, \\
\dot v(t)&=c-d u^2(t)-v(t),\\
\dot w(t)&=\epsilon[s(u(t)-r)-w(t)].
\end{aligned}
\end{equation}
Here $u$, $v$, $w$ are the state variables, $I$ is an external stimulus. $a=1$, $b=3$, $d=5$, $s=4$, $r=-1.6$, $\epsilon=1\times10^{-3}$ are some constants. Depending on the parameter $c$ the system \eqref{f8} can generate different regimes. We have chosen that parameter because it is an equivalent to parameter $a$ from FHN system. If $c > 2.2$ then system oscillates, while $0< c <2.2 $ system is excitable. In this model $u$ describes the dynamics of the membrane potential, while $v$ and $w$ illustrate how the sodium-potassium pump works. Since the rate of changing $w$ is determined by $0<\epsilon\ll1$, $v$ describes the dynamics of the slow potassium current, while $w$ describes the dynamics of the fast sodium current. 

As before, introduce new time $\tilde t=\delta t$ to make the system \eqref{f8} solution oscillate with the same frequency as gamma rhythm $x(t)$. Again we choose $\delta=325$. Adding the low-frequency delta rhythm $y(t)$ to the first equation of \eqref{f8} as an external stimulus to affect the threshold $c$, and adding the ``neuronal'' unbiased Gaussian white noise $\xi(t)$, we present the system \eqref{f8} equation as (with omitted tilde):

\begin{equation}\label{f9}
\begin{aligned}
\dot u(t)&=\delta[v(t)-au^3(t)+bu^2(t)-\\ 
&~~~~~~~~~~~~~~-w(t)+\xi(t) + y(t+\tau)], \\
\dot v(t)&=\delta[c-d u^2(t)-v(t)],\\
\dot w(t)&=\delta\epsilon[s(u(t)-r)-w(t)].
\end{aligned}
\end{equation}

\begin{figure}
\flushleft
\begin{minipage}[h]{1\linewidth}
\center{\includegraphics[width=1\linewidth]{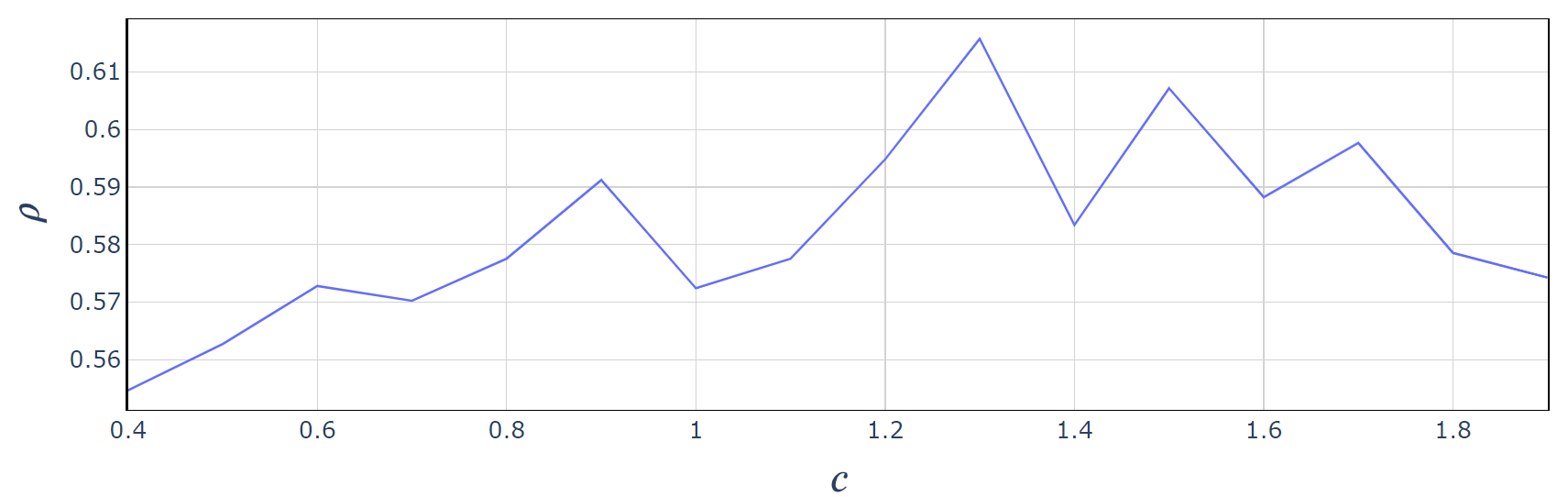} \\ (a)}
\end{minipage}
\vfill
\begin{minipage}[h]{1\linewidth}
\center{\includegraphics[width=1\linewidth]{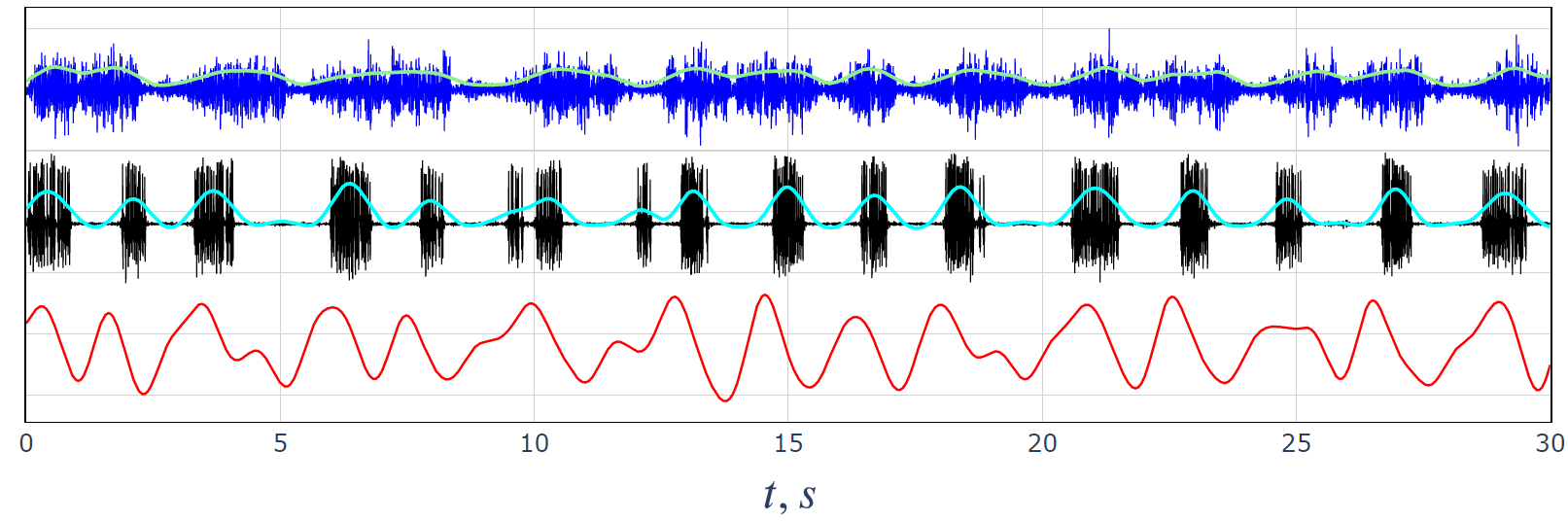} \\ (b)}
\end{minipage}
\caption{The comparison of experimental and generated by HR model signals. (a): experimental $e_n$ and HR model $e_{sim}$ envelopes correlation dependence on the parameter $c$; (b): dynamics of the HR system membrane potential $u(t)$ \eqref{f9} (marked by black color) and its envelope (marked by cyan color) with comparison to the high-frequency gamma rhythm $x(t)$ (marked by blue color) and its envelope (marked by green color). The red color marks the low-frequency delta rhythm $y(t)$. System parameters: $a=1$, $b=3$, $c=1.3$, $d=5$, $s=4$, $r=-1.6$ $\epsilon=1\times10^{-3}$, $\delta=325$, $\tau=-0.456$. The initial conditions are zeros.}
\label{fig6}
\end{figure}

Varying the value of parameter $r$ between $0.4$ and $2$ with step $0.1$ we calculate the Pearson correlation coefficient between high-frequency signal envelope $e_n$ and simulated signal envelope $e_{sim}$. Result are presented on the Fig.~\ref{fig6}~(a). The maximal value of correlation is $0.62$, which corresponds to the parameter value $c=1.3$.

The simulation results of the HR system \eqref{f9} dynamics are presented in Fig.~\ref{fig6}~(b). As before, the low frequency signal $y(t)$ affects the HR system \eqref{f7} threshold: the system exhibits self-sustained periodic firing, when $y(t)$ value becomes high enough, while there is the oscillation death when $y(t)$ value is low. The periods of HR system \eqref{f7} firing also coincide with the periods of gamma rhythm $x(t)$ activation. Thus, we can also use HR model to simulate the gamma rhythm.

\section{Conclusion}\label{sec:concl}
In this paper we have studied the dependencies between fast gamma rhythm and delta rhythm. We have performed the signal processing of the ECoG recordings of the simple Wistar rats and have shown that the delta rhythm modulates the gamma rhythm with some time delay.  The increasing of low-frequency signal value leads to the emergence of gamma oscillations, while its decreasing leads to the oscillation death of gamma rhythm. This means that there is a threshold value which determines the dynamics of gamma rhythm.

Also we have considered two neuron models, namely FHN and HR, have tuned its parameters and have shown that they can modulate high-frequency signal like gamma rhythm where delta rhythm serves as the system input. Now FHN model has more appropriate results, but since HR model has six parameters, we assumed that simultaneous variation several of them should give much better correlation value.

 \bibliography{ifacconf}     
 
\end{document}